\documentclass[letterpaper, 10 pt, conference]{ieeeconf}  

\IEEEoverridecommandlockouts                              

\usepackage{bchart,caption,subcaption,pgfplots}

\def\Rset{\mathbb{R}}

\newcommand{\comment}[1]{}
\def\ttt{^{\mbox{\tiny\sf T}}}
\def\diag{{\mathrm {\tt diag}}}
\def\bfo{{\mathbf 1}}
\def\hmu{\hat{\mu}}
\def\tm{\tilde{m}}
\def\tmu{\tilde{\mu}}

\usepackage{graphics} 
\usepackage{times} 
\usepackage[scientific-notation=true]{siunitx}
\usepackage{mathtools,amsmath,amssymb}
\usepackage[colorlinks=true]{hyperref}

\newtheorem{theorem}{Theorem}
\newtheorem{proposition}[theorem]{Proposition}
\newtheorem{lemma}{Lemma}

\newtheorem{definition}{Definition}

\definecolor{mygreen}{rgb}{0.05,0.55,0.1}

\title{\LARGE \bf
	Modelling and control of Mendelian and maternal inheritance\\
	for biological control of dengue vectors*}

\author{Pastor E.\ P\'erez Estigarribia$^1$, Pierre-Alexandre Bliman$^{2,\star}$ and
	Christian E.\ Schaerer$^1$
	\thanks{*All authors acknowledge support of STIC AmSud program (project 20-STIC-05), and P.E.P.E.\ and C.S.\ of FEEI-CONACYT-PROCIENCIA program. $^\star$Corresponding author}
	\thanks{$^{1}$P.E.\ P\'erez Estigarribia and C.E.\ Schaerer are with Polytechnic School, National University of Asunci\'on, P.O.\ Box 2111 SL, San Lorenzo, Paraguay {\tt\small \{peperez.estigarribia,cschaer\}@pol.una.py}}
	\thanks{$^2$ Pierre-Alexandre Bliman is with Sorbonne Universit\'e, Universit\'e Paris-Diderot SPC, Inria, CNRS, Laboratoire Jacques-Louis Lions, \'equipe MAMBA, F-75005 Paris, France {\tt\small pierre-alexandre.bliman@inria.fr}}
}

\begin{document}

	\maketitle
	\thispagestyle{empty}
	\pagestyle{empty}

	\begin{abstract}
		
		Mosquitoes are vectors of viral diseases  with epidemic potential in many regions of the world, and in absence of vaccines or therapies, their control is the main alternative.
		Chemical control through insecticides has been one of the conventional strategies, but induces insecticide resistance, which may affect other insects and cause ecological damage.
		Biological control, through the release of mosquitoes infected by the maternally inherited bacterium {\em Wolbachia}, which inhibits their vector competence, has been proposed as an alternative.
		The effects of both techniques may be intermingled in practice:
		prior insecticide spraying may debilitate wild population, so facilitating subsequent invasion by the bacterium; but the latter may also be hindered by the release of susceptible mosquitoes in an environment where the wild population became resistant, as a result of preexisting undesired exposition to insecticide.
		To tackle such situations, we propose here a unifying model allowing to account for the cross effects of both control techniques, and based on the latter, design release strategies able to infect a wild population.
		The latter are feedback laws, whose stabilizing properties are studied.
	\end{abstract}

	\section{Introduction}
	\label{se1}
	
	Due to 
	resistance evolution \cite{Brown:1986aa,Hemingway:2000aa,Schechtman:2015aa,Levick:2017aa} and potential ecological damages \cite{Oosthoek:2013aa}, progress has been made in the last fifteen years in the development of strategies alternative to the chemical control of vectors, ranging from biological control to genetic modification \cite{McGraw:2013aa,Kean:2015aa,Achee:2019aa}. These control techniques may have the purpose of suppressing a population, 
	or replacing it by mosquitoes with reduced or null vector competence \cite{Walker:2011aa,Alphey:2014aa,Leftwich:2016aa}.
	One of the new promising strategies is the use of {\em Wolbachia}, an intracellular bacterium passed in the insect from mother to  offspring 
	that, depending on the strain, can reduce vector competence of {\em Aedes} species relative to arboviruses transmissible to humans \cite{McMeniman:2009aa,Hoffmann:2011aa}.
	Mathematical models for the release of mosquitoes infected by {\em Wolbachia} have been proposed, see {\em e.g.}\ \cite{Keeling:2003aa,Farkas:2010aa,Zheng:2014aa,Yakob:2017aa,Xue:2017aa,Campo-Duarte:2017aa,
		Bliman:2018aa,Almeida:2019aa}, and \cite{Hughes:2013aa,Koiller:2014aa,Ndii:2015aa} in the context of dengue epidemic.

	The use of chemical control  has been mentioned as a way to facilitate the incorporation of {\em Wolbachia} into a population \cite{Hoffmann:2013aa}.
	On the other hand, it has been reported that undesired exposition to insecticide may weaken the action of released susceptible mosquitoes against resistant wild population \cite{Garcia:2012aa,Garcia:2017aa,Azambuja-Garcia:2019aa}.
	The purpose of this note is to provide a model capable to describe such complex situations and offer a framework to design feedback control laws aiming at spreading {\em Wolbachia} infection among mosquito population.
	
	
	A 12-dimensional controlled model with pre-reproductive and reproductive life phases is provided in Section \ref{se2}, accounting for Mendelian inheritance of insecticide resistance, as well as maternal transmission of {\em Wolbachia} infection.
	Assuming that the larval phase is significantly faster than the adult phase, the model is then simplified by slow manifold theory into the 6-dimensional model \eqref{eq0}, whose qualitative properties are studied.
	This extends the model of Mendelian inheritance in \cite{Perez:2020aa}, following \cite{Langemann:2013aa}, which constitutes a model of diploid population with setting of two alleles in a single locus.
	Section \ref{se3} provides useful balance equations.
	Number and stability of the equilibria of \eqref{eq0} in absence of release is the subject of Section \ref{se4}.
	Results of stabilization by state-feedback are then given in Section \ref{se5}, applying ideas from \cite{Bliman:2020aa}.
	Under adequate assumptions, they allow to infect the wild population even in presence of insecticide in the environment, yielding installation of the resistant infected population.
	Illustrative simulations are shown in Section \ref{se6}.
	For sake of space, only hints of proofs are provided.
	
	\section{Controlled models}
	\label{se2}
	
	\subsection{Preliminaries}
	
	\subsubsection{Notations}
	
	Generally speaking, in the sequel the capital letters $U, W$ refer to uninfected ($U$) and {\em Wolbachia}-infected ($W$) populations; while the small letter $j=r,s$ refers to the two alleles ({\em resistant} and {\em susceptible}) and the number $i=1,2,3$  to the three genotypes.
	By convention, $i=1$, resp.\ $i=2$, resp.\ $i=3$ corresponds to genotype $(r,r)$, resp.\ $(r,s)$, resp.\ $(s,s)$.
	In the notations adopted below, the reference to infective status is put in exponent, while the information relative to genotypic/allelic status is put in index.
	
	We define the vectors
	$e^U := \begin{pmatrix}
	1 & 0
	\end{pmatrix}\ttt$,
	$e^W := \begin{pmatrix}
	0 & 1
	\end{pmatrix}\ttt$,
	$e_1 := \begin{pmatrix}
	1 & 0 & 0
	\end{pmatrix}\ttt$,
	$e_2 := \begin{pmatrix}
	0 & 1 & 0
	\end{pmatrix}\ttt$,
	$e_3 := \begin{pmatrix}
	0 & 0 & 1
	\end{pmatrix}\ttt$,
	$\bfo_2 := \begin{pmatrix} 1 & 1 \end{pmatrix}\ttt$,
	$\bfo_3 := \begin{pmatrix} 1 & 1 & 1 \end{pmatrix}\ttt$,
	$\bfo_6 := \bfo_2\otimes\bfo_3$, where $\otimes$ denotes the Kronecker product.
	Kronecker delta is defined as usual: for any $a,b$, $\delta_a^b=1$ if $a=b$, $\delta_a^b=0$ otherwise.
	Last, for any $z\in\Rset$, let $|z|_+ := \max\{z;0\}$.
	
	\subsubsection{State variables}
	\label{se212}
	
	\begin{figure*}[h]
		\begin{gather}
		\label{eq778}
		G_1 	=
		\begin{pmatrix}
		1 & 1/2 & 0\\ 1/2 & 1/4 & 0\\ 0 & 0 & 0
		\end{pmatrix},\quad
		G_2
		= \begin{pmatrix}
		0 & 1/2 & 1\\ 1/2 & 1/2 & 1/2 \\ 1 & 1/2 & 0
		\end{pmatrix},\quad
		G_3
		= \begin{pmatrix}
		0 & 0 & 0\\ 0 & 1/4 & 1/2 \\ 0 &1/2 & 1
		\end{pmatrix}\\
		\label{eq500-501}
		\alpha^\eta_i(A) := \frac{1}{|A|} A\ttt (G^\eta\otimes G_i) A,
		\qquad
		\alpha(A) := \sum_{\eta=U,W}\sum_{i=1,2,3} \alpha^\eta_i(A) (e^\eta\otimes e_i)\\
		\label{eq7-10}
		L^\eta_i = \frac{\omega^\eta \alpha^\eta_i(A)}{\hmu^\eta_i(b^*(\alpha(A))) + \nu}, \qquad
		b^*(A) = \sum_{i=1}^3\sum_{\eta=U,W} \frac{\omega^\eta A^\eta_i}{\hmu^\eta_i(b^*(A)) + \nu}
		\end{gather}
		\hrule
	\end{figure*}
	
	For $\eta\in\{U,W\}$ and $i\in\{1,2,3\}$, denote $A^\eta_i\in\Rset_+$ the density of adults of genotype $i$ uninfected ($\eta=U$) or {\em Wolbachia}-infected ($\eta=W$).
	Let $A := \begin{pmatrix} A^U_1 & A^U_2 & A^U_3 & A^W_1 & A^W_2 & A^W_3 \end{pmatrix}\ttt$,
	$A^\eta := \begin{pmatrix} A^\eta_1 & A^\eta_2 & A^\eta_3\end{pmatrix}\ttt$ for $\eta =U,W$, and $A_i := \begin{pmatrix} A^U_i & A^W_i \end{pmatrix}\ttt$ for $i=1,2,3$.
	One shows easily that $A = \sum_{i=1}^3 (A_i\otimes e_i) = \sum_{\eta=U,W} (e^\eta\otimes A^\eta)$.
	We will use the $L^1$ norm of these vectors, denoted $|A| := \bfo_6\ttt A$, $|A^\eta| := \bfo_3\ttt A^\eta$, $|A_i| := \bfo_2\ttt A_i$.
	
	We will also consider the densities of alleles in the uninfected and infected populations, namely
	$A^\eta_r := A^\eta_1+\frac{1}{2} A^\eta_2$ and $A^\eta_s := A^\eta_3+\frac{1}{2} A^\eta_2$, for any $\eta=U,W$.
	One then has\footnote{By slight abuse of notations, one denotes indifferently in index the genotypes (using letter $i=1,2,3$) or the alleles (using letter $j=r,s$).} the {\em vectorial} identities: $A_r = A_1+\frac{1}{2}A_2$, $A_s = A_3+\frac{1}{2}A_2$.
	We will in general write such formulas as $A_j = A_i+\frac{1}{2}A_2$, {\em adopting the convention that $i=1$ (resp.\ $i=3$) whenever $j=r$ (resp.\ $j=s$)}.
	Coherently with the previous notations, define the norms $|A_j| := A^U_j+A^W_j$, $j=r,s$.
	
	Similar notations are used for the early phase densities $L^\eta_i$.
	
	We make the following qualitative definitions.
	\begin{definition}[Monomorphic and polymorphic states]
		Any point $A\in\Rset_+^6$ is called a {\em monomorphic state} if it contains only one allele, {\em i.e.}\ $A^\eta_1>0=A^\eta_2=A^\eta_3$ for every $\eta=U,W$, or $A^\eta_1=A^\eta_2=0<A^\eta_3$ for every $\eta=U,W$.
		Non-monomorphic points are called {\em polymorphic}.
	\end{definition}
	
	\begin{definition}[Homogeneous and heterogeneous states]
		Any point $A\in\Rset_+^6$ is called a {\em homogeneous state} if it contains only uninfected populations, that is $A^W_i=0$, $i=1,2,3$; or if it contains only {\em Wolbachia} infected populations, that is $A^U_i=0$, $i=1,2,3$.
		Non-homogeneous points are called {\em heterogeneous}.
	\end{definition}
	
	\subsubsection{Inheritance modelling}
	
	In order to deal with maternal inheritance of {\em Wolbachia} (with complete cytoplasmic incompatibility, defined later), we need the following notations:
	$\displaystyle G^U := e^U e^{U\mbox{\tiny\sf T}}= \begin{pmatrix}
	1 & 0 \\ 0 & 0
	\end{pmatrix}$,
	$\displaystyle G^W := \bfo_2 e^{W\mbox{\tiny\sf T}}= \begin{pmatrix}
	0 & 1 \\ 0 & 1
	\end{pmatrix}$.

	On the other hand, handling of the Mendelian inheritance will require
	$u_r := \begin{pmatrix}
	1 & 1/2 & 0
	\end{pmatrix}\ttt$,
	$u_s := \begin{pmatrix}
	0 & 1/2 & 1
	\end{pmatrix}\ttt$, and
	$G_1 := u_ru_r\ttt, G_2 := u_ru_s\ttt + u_su_r\ttt, G_3 := u_su_s\ttt$, that is \eqref{eq778}.
	Notice that $u_r+u_s=\bfo_3$ and $G_1+G_2+G_3 = \bfo_3\bfo_3\ttt$.

	We now define a key notion, the {\em heredity functions}, which give {\em the repartition of offspring of a population $A$, according to the distribution of genotypes and infectiousness}.
	These are scalar functions $\alpha^\eta_i\ :\ \Rset_+^6\setminus\{0_6\}\to\Rset_+$, $\eta=U,W$, $i=1,2,3$, given in \eqref{eq500-501}, which permit to form the {\em matrix heredity function} $\alpha$.
	One extends by continuity the previous definitions to $0_6$, putting 
	$\alpha(0_6) = 0_6$.
	Observe that $\alpha$ so defined is positively homogeneous of degree 1.
	
	\subsection{Complete and reduced inheritance models}
	\label{se21}
	Introducing input variables $v_{Ai}$ corresponding to releases of infected adults of genotypes 1, 2 or 3, yields the following {\em controlled} model of {\em Wolbachia} infection in presence of insecticide resistance:
	\begin{subequations}
		\label{eq1}
		\begin{gather}
		\label{eq1a}
		\dot L^\eta_i = \omega^\eta \alpha^\eta_i(A) - \hmu^\eta_i(|L|) L^\eta_i - \nu L^\eta_i \\
		\label{eq1b}
		\dot A^\eta_i = \nu L^\eta_i - \mu^\eta_i A^\eta_i + \delta_\eta^W v_{Ai},
		\end{gather}
	\end{subequations}
	$\eta=U,W$, $i=1,2,3$, where the $\alpha^\eta_i$ are defined in \eqref{eq500-501}.
	The positive constants $\omega^\eta$, resp.\ $\mu^\eta_i$, are fertility, resp.\ mortality, rates.
	The mortality rate $\hmu^\eta_i(|L|)$ in pre-reproductive phase, is an increasing function of the density $|L|$
	in this phase.
	The maturation rate $\nu$ from pre-reproductive to reproductive phase, is taken independent of genotype and infection.
	
	The functions $\alpha^\eta_i$ account for the inheritance mechanisms.
	Considering all possible crosses given a random mating, the expected frequency of two allele combinations in a diploid population
	is obtained from Punnett Square \cite{Edwards:2012aa}.
	This is captured by the matrices $G_i$, $i=1,2,3$.
	On the other hand, Wolbachia induces {\em cytoplasmic incompatibility} (CI): when an uninfected female is inseminated by an infected male, the mating leads to sterile eggs.
	This crossing effect is grasped by the matrices $G^\eta$, $\eta=U,W$ (CI is complete here: no viable offspring hatch from such an encounter, see \cite{Perez:2020ab} for modelling of incomplete CI).

	
	The pre-adult phase being fast relatively to the adult one, one approximates the system by singular perturbation.
	Consider instead of \eqref{eq1a} the algebraic formula
	$\omega^\eta \alpha^\eta_i(A) - \hmu^\eta_i(|L|) L^\eta_i - \nu L^\eta_i =0$, 
	yielding
	$L^\eta_i = \frac{\omega^\eta \alpha^\eta_i(A)}{\hmu^\eta_i(|L|) + \nu}$.
	Summing up these six expressions gives an equation in the unknown $|L|$, which by standard argument has unique solution if the functions $\hmu^\eta_i$ are increasing (see Section \ref{se22}).
	This solution is written $b^*(\alpha(A))$, for $b^*\ :\ \Rset_+^6\to\Rset_+$ defined implicitly in \eqref{eq7-10}.
	The components $L^\eta_i$ may then be expressed with respect to the $A^\eta_i$, see \eqref{eq7-10}.
	Putting these expressions in \eqref{eq1b} yields
	\begin{subequations}
		\label{eq10}
		\begin{equation}
		\label{eq10a}
		\dot A^\eta_i = m^\eta_i (b^*(\alpha(A)))\, \alpha^\eta_i (A) - \mu^\eta_i A^\eta_i + \delta_\eta^W v_{Ai},
		\end{equation}
		where $b^*$ is defined in \eqref{eq7-10} and, for any $b\in\Rset_+$,
		\begin{equation}
		\label{eq10b}
		m^\eta_i (b) := \frac{\nu \omega^\eta}{\hmu^\eta_i(b) + \nu}.
		\end{equation}
	\end{subequations}
	\begin{figure*}[h]
		\begin{subequations}
			\label{eq00}
			\begin{eqnarray}
			\label{eq00a}
			\dot A^U_i
			& = &
			\frac{m^U_i(b^*(\alpha(A)))}{|A|} \left(
			A^U_i+\frac{1}{2}A^U_2
			\right) \left(
			A^U_i+\frac{1}{2}A^U_2
			\right)
			- \mu^U_iA^U_i,\quad i=1,3 \\
			\dot A^U_2
			& = &
			\label{eq00b}
			2 \frac{m^U_2(b^*(\alpha(A)))}{|A|} \left(
			A^U_1+\frac{1}{2}A^U_2
			\right) \left(
			A^U_3+\frac{1}{2}A^U_2
			\right)
			- \mu^U_2A^U_2\\
			\label{eq00c}
			\dot A^W_i
			& = &
			\frac{m^W_i(b^*(\alpha(A)))}{|A|} \left(
			A^U_i+\frac{1}{2}A^U_2 + A^W_i+\frac{1}{2}A^W_2
			\right) \left(
			A^W_i+\frac{1}{2}A^W_2
			\right)
			- \mu^W_iA^W_i + v_{Ai}(t),\quad i=1,3\\
			\dot A^W_2
			& = &
			\nonumber
			\frac{m^W_2(b^*(\alpha(A)))}{|A|} \left(
			A^U_3+\frac{1}{2}A^U_2 + A^W_3+\frac{1}{2}A^W_2
			\right) \left(
			A^W_1+\frac{1}{2}A^W_2
			\right)\\
			&  &
			\label{eq00e}
			+ \frac{m^W_2(b^*(\alpha(A)))}{|A|} \left(
			A^U_1+\frac{1}{2}A^U_2 + A^W_1+\frac{1}{2}A^W_2
			\right) \left(
			A^W_3+\frac{1}{2}A^W_2
			\right)
			- \mu^W_2A^W_2 + v_{A2}(t)
			\end{eqnarray}
		\end{subequations}
		\hrule
	\end{figure*}
	The nonlinear controlled density-dependent inheritance system \eqref{eq10} is developed in \eqref{eq00}, and writes compactly as:
	\begin{subequations}
		\label{eq0}
		\begin{gather}
		\label{eq0a}
		\dot A = m(b^*(\alpha(A)))\, \alpha(A) - \mu A+ \begin{pmatrix} 0_3 \\ v_A(t) \end{pmatrix},\\
		\label{eq0b}
		m(b):=\diag\{ m^\eta_i (b) \},\quad
		\mu :=\diag\{\mu^\eta_i\}
		\end{gather}
	\end{subequations}
	

	
	
	\subsection{Assumptions on the dynamical system \eqref{eq0}}
	\label{se22}
{\em Wolbachia} infection induces fitness reduction~\cite{Walker:2011aa,Hoffmann:2014aa,Koiller:2014aa}, and in presence of insecticide, resistant mosquitoes have larger fitness than susceptible ones.
We thus posit that,
	for any $\eta=U,W$, $i,i'=1,2,3$,
	\begin{itemize}
		\item
		$\mu^\eta_i\ :\ \Rset_+\to\Rset_+$ non-decreasing; $\hmu^\eta_i\ :\ \Rset_+\to\Rset_+$ increasing and unbounded
		\item
		$i < i'$ implies $\hmu^\eta_i \leq \hmu^\eta_{i'}$ and $\mu^\eta_i \leq \mu^\eta_{i'}$
		\item
		$\hmu^U_i \leq \hmu^W_i$ and $\mu^U_i \leq \mu^W_i$
	\end{itemize}
	Moreover, we assume that some of the previous inequalities are {\em strict} (see details in \cite{Perez:2020aa,Perez:2020ab}), by assuming that
	\begin{itemize}
		\item
		$\mu^U_1 < \mu^W_2$.
	\end{itemize}
	
	One deduces easily from the previous assumptions that
	\begin{itemize}
		\item
		$m^\eta_i :\ \Rset_+\to\Rset_+$ decreasing with limit zero
		\item
		$i > i'$ implies $m^\eta_i > m^\eta_{i'}$
		\item
		$m^U_i > m^W_i$
	\end{itemize}
	and in particular, the definition of $b^*$ in \eqref{eq7-10} is meaningful.
	
	\subsection{Well-posedness and qualitative properties}
	
	We assume in the sequel that the control input $v_A$ is locally integrable and almost everywhere positive.
	Showing the well-posed of system \eqref{eq0} then presents no specific difficulty.
	
	We first establish that any genotype once present may only disappear in infinite time, whatever the control input.
	\begin{theorem}[Polymorphic and heterogeneous trajectories]
		\label{th15}
		Whatever the (nonnegative-valued) input signal $v_A$, all trajectories of system \eqref{eq0} fulfil the following properties. 
		\begin{enumerate}
			\item
			\label{15i}
			For any trajectory such that $A^\eta_i(0)>0$ for some $\eta=U,W$, $i=1,2,3$, one has $A^\eta_i(t)>0$ for any $t\geq 0$.
			\item
			\label{15ii}
			Any trajectory originating from monomorphic (resp.\ homogeneous) state remains monomorphic (resp.\ homogeneous) for any $t\geq 0$ if no other genotype (resp.\ no population with other infection status) is introduced.
			\item
			\label{15iii}
			Any trajectory originating from polymorphic (resp.\ heterogeneous) state remains polymorphic (resp.\ heterogeneous) for any $t\geq 0$.
			\hfil\hfil $\blacksquare$
		\end{enumerate}
	\end{theorem}
	As a consequence, 
	one may talk about {\em homogeneous} or {\em heterogeneous trajectories}, and similarly about {\em monomorphic} or {\em polymorphic trajectories}.
	We now study boundedness.
	
	\begin{theorem}[Trajectory boundedness]
		\label{th55}
		Assume the input control $v_A$ uniformly bounded on $[0,+\infty)$.
		Then all trajectories of \eqref{eq0} are uniformly ultimately bounded.
		\hfil\hfil $\blacksquare$
	\end{theorem}
	
	The proof comes from the inequality
	$\frac{d|A|}{dt} \leq \left(
	m^U_1(b^*(\alpha(A))) - \mu^U_1
	\right) |A| +\ \|v_A\|_{L^\infty}$,
	and the fact that $m^U_1(b)$ is decreasing and vanishes at infinity.
	
	The last result unveils 
	some mixing properties, 
	characteristic of the underlying genetic mechanisms involved.
	\begin{theorem}[Genotypic properties]
		\label{th56}
		For any nonnegative-valued input signal, the trajectories of system \eqref{eq0} fulfil the following properties. 
		\begin{enumerate}
			\item
			If both alleles are present at $t=0$ in the uninfected (resp.\ infected) population, then all genotypes are present in the uninfected (resp.\ infected) population for any $t>0$.\label{gp1}
			\item
			If some allele is present at $t=0$ in the uninfected population and the other one in the infected, then all genotypes are present in the {\em infected} population for any $t>0$.\label{gp2}
			\item
			If only the allele $j\in\{r,s\}$ is present at $t=0$ in the {\em uninfected} population ({\em i.e.}\ $A^U_1(0)>0=A^U_2(0)=A^U_3(0)$ if $j=r$, or $A^U_1(0)=A^U_2(0)=0<A^U_3(0)$ if $j=s$), then the same holds true for any $t\geq 0$.\label{gp3}
			\hfil\hfil $\blacksquare$
		\end{enumerate}
	\end{theorem}
	
	The proof of Theorem \rm\ref{th56} uses centrally results from Theorem \ref{th15} and Lemma \ref{le34}.

	\section{Balance equations}
	\label{se3}
	
	Summing equations in \eqref{eq10} yields interesting balance equations.
	We provide an allelic description of the evolution in Section \ref{se31}, and uninfected/infected balance equations in Section \ref{se32}.
	None of them 
	forms a replicator equation \cite{Hofbauer:1998aa}.
	
	
	\subsection{Evolution at allelic level}
	\label{se31}
	
	We aim here at a description in terms of the 4 allelic variables $A^\eta_j$, $\eta=U,W$, $j=r,s$.
	From \eqref{eq10a} and with the definitions in Section \ref{se212}, one gets by summation
	\begin{equation}
	\label{eq110}
	\hspace{-.25cm}
	\dot A^\eta_j = \tm^\eta_j (A)\,
	\left(
	\alpha^\eta_i(A)+\frac{1}{2} \alpha^\eta_2(A)
	\right)
	- \tmu^\eta_j(A) A^\eta_j + \delta_\eta^W v_{Ai},
	\end{equation}
	$\eta=U,W$, $j=r,s$, where the {\em mean allelic recruitment and mortality rates} $\tm^\eta_j(A),\tmu^\eta_j$ are defined in \eqref{eq5100}.
	\begin{figure*}[h]
		\begin{gather}
		\label{eq5100}
		\tm^\eta_j(A) := \frac{\alpha^\eta_i(A) m^\eta_i(b^*(\alpha(A))) +\frac{1}{2} \alpha^\eta_2(A) m^\eta_2(b^*(\alpha(A)))}{\alpha^\eta_i(A)+\frac{1}{2} \alpha^\eta_2(A)},\quad
		\tmu^\eta_j(A) :=
		\frac{A^\eta_i \mu^\eta_i +\frac{1}{2} A^\eta_2 \mu^\eta_2}{A^\eta_i+\frac{1}{2} A^\eta_2},\quad
		\eta=U,W,\ j=r,s\\
		\label{eq16-117}
		\hspace{-.2cm}
		\frac{d|A^\eta|}{dt}
		= \tm^\eta(t) \sum_{i=1}^3 \alpha^\eta_i (A) - \tmu^\eta(t) \sum_{i=1}^3 A^\eta_i +\ \delta_\eta^W |v_A|,\
		\tm^\eta(t) := \frac{\sum_{i=1}^3 m^\eta_i (b^*(\alpha(A)))\alpha^\eta_i (A)}{\sum_{i=1}^3 \alpha^\eta_i (A)},\
		\tmu^\eta(t) := \frac{\sum_{i=1}^3 \mu^\eta_iA^\eta_i}{\sum_{i=1}^3 A^\eta_i}\\
		\label{eq111}
		\hspace{-.2cm}
		\frac{m^W_i(b^*(\alpha(A^{U**}_j (e^U \otimes e_i) + A^{W**}_j (e^W \otimes e_i))))}{\mu^W_i} = 1,\
		\frac{A^{W**}_j}{A^{U**}_j} = \frac{m^U_i(b^*(\alpha(A^{U**}_j (e^U \otimes e_i) + A^{W**}_j (e^W \otimes e_i))))}{\mu^U_i}-1
		\end{gather}
		\hrule
	\end{figure*}
	(The convention: $i=1$ for $j=r$, $i=3$ for $j=s$ is used.)
	Using Lemma \ref{le1}, one may express system \eqref{eq110} as
	\begin{subequations}
		\label{eq5102}
		\begin{gather}
		\label{eq5102a}
		\dot A^U_j = \left(
		\tm^U_j (t)\frac{|A^U|}{|A|} - \tmu^U_j(t)
		\right) A^U_j,\\
		\label{eq5102b}
		\dot A^W_j = \frac{1}{2} \tm^W_j (t)\left(
		A^W_j + \frac{|A_j| |A^W|}{|A|}
		\right) - \tmu^W_j(t) A^W_j + v_{Aj},
		\end{gather}
	\end{subequations}
	$ j=r,s$, or in expanded form:
	\begin{subequations}
		\label{eq5103}
		\begin{gather}
		\label{eq5103a}
		\dot A^U_r = \left(
		\tm^U_r (t)\frac{|A^U|}{|A|} - \tmu^U_r(t)
		\right) A^U_r\\
		\label{eq5103b}
		\dot A^U_s = \left(
		\tm^U_s (t)\frac{|A^U|}{|A|} - \tmu^U_s(t)
		\right) A^U_s\\
		\label{eq5103c}
		\dot A^W_r = \frac{1}{2} \tm^W_r (t)\left(
		A^W_r + \frac{|A_r| |A^W|}{|A|}
		\right) - \tmu^W_r(t) A^W_r + v_{Ar}\\
		\label{eq5103d}
		\dot A^W_s = \frac{1}{2} \tm^W_s (t)\left(
		A^W_s+ \frac{|A_s| |A^W|}{|A|}
		\right) - \tmu^W_s(t) A^W_s + v_{As}
		\end{gather}
	\end{subequations}
	
	Formally, one may interpret \eqref{eq5102} as describing the infection of two populations of alleles.
	But the situation is more intricate, as the coefficients appearing are not merely functions of the $A^\eta_j$, see  \eqref{eq5100}.
	However, they fulfil the following useful properties, for any $A\in\Rset_+^6\setminus\{0_6\}$:
	\begin{subequations}
		\label{eq620}
		\begin{multline}
		\label{eq620a}
		m^\eta_3(b^*(\alpha(A))) \leq \tm^\eta_s(A) \leq m^\eta_2(b^*(\alpha(A)))\\
		\leq \tm^\eta_r(A) \leq m^\eta_1(b^*(\alpha(A))),
		\end{multline}
		\vspace{-.8cm}
		\begin{gather}
		\label{eq620b}
		\mu^\eta_1 \leq \tmu^\eta_r(A) \leq \mu^\eta_2 \leq \tmu^\eta_s(A) \leq \mu^\eta_3,\\
		\label{eq620c}
		\tm^W_j(A) \leq \tm^U_j(A),\quad \tmu^U_j(A) \leq \tmu^W_j(A).
		\end{gather}
	\end{subequations}
	
	
	\subsection{Uninfected/infected balance equations}
	\label{se32}
	
	For $\eta = U,W$, the evolution of $|A^\eta| = A^\eta_1+A^\eta_2+A^\eta_3$ obeys equation \eqref{eq16-117}, where we put by definition $ |v_A|:=\sum_{i=1}^3 v_{Ai}$.
	This may be expressed as
	$\frac{d|A^\eta|}{dt}
	= \tm^\eta(t) \alpha^\eta \left(
	A
	\right) - \tmu^\eta(t) |A^\eta| +\ \delta_\eta^W |v_A|$, 
	$\eta=U,W$, or in developed form, as
	\begin{subequations}
		\label{eq1118}
		\begin{gather}
		\label{eq1118a}
		\frac{d|A^U|}{dt} = \left(
		\tm^U(t) \frac{|A^U|}
		{|A^U|+|A^W|} - \tmu^U(t)
		\right) |A^U|\\
		\label{eq1118b}
		\frac{d|A^W|}{dt} = (\tm^W(t) - \tmu^W(t)) |A^W| +\ |v_A|
		\end{gather}
	\end{subequations}
	
	Similarly to \eqref{eq5103}, equation \eqref{eq1118} describes an evolution which is only apparently independent of the infection status, as the latter is involved in the mean recruitment and mortality rates $\tm^\eta, \tmu^\eta$, $\eta=U,W$, defined in \eqref{eq16-117}.
	One checks easily from the assumptions that $\tm^U \geq \tm^W$, $\tmu^U\leq \tmu^W$, and
	$m^\eta_3 \leq \tm^\eta \leq m^\eta_1$, $\mu^\eta_1 \leq \tmu^\eta \leq \mu^\eta_3$, $\eta=U,W$.
	
	\section{Equilibria of the uncontrolled system}
	\label{se4}
	
	We study here the number and properties of the equilibrium points of the uncontrolled system \eqref{eq0}.
	
	\subsection{Existence of equilibrium points}
	\label{se41}
	
	First is determined the number and type of the equilibria.
	
	\begin{theorem}[Equilibria of \eqref{eq0} with $v_A\equiv 0$]
		\label{th8}
		Apart from the {\em extinction equilibrium} $0_6$,
		the equilibrium points of the uncontrolled
		system \eqref{eq0} fulfil the following properties.
		\begin{itemize}
			\item
			There are {\em at most six monomorphic equilibrium points}:
			
			- at most four {\em monomorphic, homogeneous, equilibria}, equal to the vectors
			$A^{\eta *}_j (e^\eta\otimes e_i)$, $\eta=U,W$, $j=r,s$,
			for $A^{\eta *}_j$ unique positive solution of the scalar equation
			$m^\eta_i \left(b^*(A^{\eta *}_j (e^\eta\otimes e_i))\right)
			= \mu^\eta_i$;
			
			- at most two {\em monomorphic, heterogeneous, coexistence equilibria}
			$A^{U**}_j (e^U \otimes e_i) + A^{W**}_j (e^W \otimes e_i)$, $j=r,s$,
			for $(A^{U**}_j, A^{W**}_j)$ unique positive solution of \eqref{eq111}.
			\item
			If any, the {\em polymorphic equilibria} fulfil
			\begin{equation}
			\label{eq104}
			A^U_3, A^W_1, A^W_2, A^W_3>0,\qquad A^U_1=A^U_2=0.
			\end{equation}
		\end{itemize}
		By convention, $i=1$ (resp.\ $i=3$) when $j=r$ (resp.\ $j=s$) in the statement.
		\hfil\hfil $\blacksquare$
	\end{theorem}
	
	Theorem \ref{th8} completely characterizes the monomorphic equilibria.
	A {\em monomorphic homogeneous equilibria} $A^{\eta *}_j (e^\eta\otimes e_i)$ distinct from extinction equilibrium exists {\em iff}
	\begin{equation}
	\label{eq101}
	m^\eta_i (0) > \mu^\eta_i.
	\end{equation}
	This condition expresses that the recruitment rate of emerging population is larger than the mortality rate, {\em i.e.}\ that the corresponding homozygous
	homogeneous population is viable for certain population level ---which is then unique and plays the role of a carrying capacity.
	
	Similarly, a nonzero {\em monomorphic heterogeneous equilibrium} $A^{U**}_j (e^U \otimes e_i) + A^{W**}_j (e^W \otimes e_i)$ exists {\em iff} \eqref{eq101} holds for $\eta=W$, and then the values of $b^*(\alpha(A^{U**}_j (e^U \otimes e_i) + A^{W**}_j (e^W \otimes e_i)))$ and of the ratio $\frac{A^{W**}_j}{A^{U**}_j}$ are uniquely determined by \eqref{eq111}.
	There is thus at most one such equilibrium for each allele.
	Notice that if \eqref{eq101} holds for $\eta=W$, it also holds for $\eta=U$, due to assumptions in Section \ref{se22}.
	
	The result concerning the {\em polymorphic equilibria} is partial: it establishes the possible general form of such a point, but does not decide about existence or uniqueness.
	
	Last, the equilibrium points of \eqref{eq0} and \eqref{eq1} being in one-to-one correspondence, Theorem \ref{th8} also holds for \eqref{eq1}.
	
	{\em Hint of proof:}
	Monomorphic (homogeneous or heterogeneous) equilibria present no difficulty.
	For any polymorphic equilibrium $A$, show first that one of the two values $A^U_r, A^U_s$ is nonzero; otherwise $\frac{\tmu^U_r}{\tm^U_r} = \frac{\tmu^U_s}{\tm^U_s}$ and, by virtue of the strict inequality assumption in Section \ref{se22}, $A^U_1=A^U_3=0>A^U_2$, which is impossible at equilibrium, see Lemma \ref{le34}.
	Assuming then $A^U_s=0$, polymorphism implies $A^W_i>0$, $i=1,2,3$, by Lemma \ref{le34}, and contradiction comes from the fact that one has $\frac{\tmu^W_r}{\tm^W_r} = \frac{\tmu^W_s}{\tm^W_s}$ at the same time.
	\hfil\hfil $\square$
	
	\subsection{Stability of the equilibrium points}
	\label{se42}
	
	We now assess stability of the equilibrium points.
	
	\begin{theorem}[Stability of the equilibria of \eqref{eq0} with $v\equiv 0$]
		\label{th9}
		All possible equilibrium points of system \eqref{eq0} are unstable, except the two homogeneous resistant monomorphic equilibria $A^{\eta*}_r(e^\eta\otimes e_1)$, $\eta=U,W$, which are locally asymptotically stable if they exist.
		\hfil\hfil $\blacksquare$
	\end{theorem}
	
	{\em Hint of proof:}
	$\bullet$ Instability of the extinction equilibrium stems from the assumed viability of the resistant populations.
$\bullet$ Any trajectory departing from {\em homogeneous, polymorphic}, state converges towards the corresponding {\em homogeneous, resistant (monomorphic)}, equilibrium $A^{\eta*}_r (e^\eta\otimes e_1)$,
the latter having higher fitness than the susceptible homozygous $A^{\eta*}_s(e^\eta\otimes e_3)$: the latter are unstable.
Same argument applies to coexistence equilibria.
	$\bullet$
	Any {\em polymorphic} equilibrium fulfils \eqref{eq104}, 
	so $\frac{d}{dt} \left[
	\ln\left(
	\frac{A^U_r}{A^U_s}
	\right)\right]
	> 0$, yielding instability by integration.
	$\bullet$
	Local asymptotic stability of the homogeneous monomorphic equilibria $A^{\eta*}_r(e^\eta\otimes e_1)$, $\eta=U,W$, comes from direct inspection of the Jacobian matrices.
	This computation requires differentiation of $\alpha$ and $b^*$, see details in \cite{Perez:2020ab}.
	\hfil\hfil $\square$
	
	\section{State-feedback stabilization}
	\label{se5}
	
	
	Consider now the issue of synthesizing state-feedback laws able to drive the system from any  initial state towards the desired resistant, fully-infected, equilibrium previously denoted $A^{W*}_r(e^W\otimes e_1)$.
	We assume from now on
	${\displaystyle \lim_{c\to 0^+}} m^\eta_1(b^*(c (e^\eta\otimes e_1))) > \mu^\eta_1$, for any $\eta=U,W$.
	In other words, 
	both {\em resistant} homozygous genotypes are viable ---a condition clearly required for lasting infection. 
	As consequence, 
	among the {\em monomorphic} equilibria exhibited in Theorem \ref{th8}, at least the {\em resistant} ones $A^{\eta *}_r (e^\eta\otimes e_1)$, $\eta=U,W$, are nonzero.
	
	
	
	
	Our aim is to control the system and reach the monomorphic equilibrium
	$(A^U_r, A^U_s,A^W_r,A^W_s) = (0,0,A^{W*}_1,0)$,
	typically (but not only) departing from the other monomorphic equilibrium
	$(A^U_r, A^U_s,A^W_r,A^W_s) = (A^{U*}_1,0,0,0)$, through release of infected alleles in \eqref{eq5103d} or \eqref{eq5103d}.
	Notice that these values, corresponding to the two different homogeneous monomorphic equilibria of resistant alleles for the underlying system \eqref{eq10a}, 
	are not really state variables:
	they constitute equilibrium points for system \eqref{eq5102}, given that the time-varying coefficients $\tm^\eta_j(t), \tmu^\eta_j(t)$, $\eta=U,W$, $j=r,s$ are in fact state-dependent quantities
	fulfilling the properties \eqref{eq620}.
	
	This task is not trivial in presence of insecticide, in case where the released infected mosquitos are susceptible.
	As a matter of fact, eliminating uninfected mosquitoes requires sufficient introduction of infected mosquitoes.
	On the other hand, the presence of resistant mosquitoes naturally forces the disappearance of susceptible ones (whose fitness is lower) through competition.
	But continued introduction of susceptible may hamper and abolish this trend.
	When infection by {\em Wolbachia} is achieved through release of susceptible mosquitoes, the two objectives ---namely {\em Wolbachia} infection and onset of insecticide resistance--- are thus potentially conflicting.
	
	\subsection{Growth rate comparison and best fitness selection}
	
	The following simple result will be instrumental.
	
	\begin{proposition}[Growth rate dominance]
		\label{th11}
		Consider positive, absolutely continuous, scalar functions $y,z$ on $[0,+\infty)$.
		Assume 
		$\frac{\dot y(t)}{y(t)} - \frac{\dot z(t)}{z(t)} \geq \varepsilon > 0$
		for some $\varepsilon>0$ and a.e.\ $t\geq 0$.
		Then
		${\displaystyle \lim_{t\to +\infty}} \frac{z(t)}{y(t)} =0$, and
		${\displaystyle \lim_{t\to +\infty}} z(t) = 0$ if $y$ is bounded.
		\hfil\hfil $\blacksquare$
	\end{proposition}
	
	The mechanism exposed in the previous result is behind the process of selection of the best fit population in a homogeneous population.
	It is applied in Proposition \ref{th14}.
	
	\begin{proposition}[Asymptotic resistance amongst uninfected]
		\label{th14}
		For any initial state $A(0)$ containing uninfected of different genotypes, consider the solution $A$ of \eqref{eq0}.
		Then $|A^U|$ is uniformly bounded along time,
		${\displaystyle \lim_{t\to +\infty}} \frac{A^U_1}{|A^U|} = 1$,
		and
		${\displaystyle \lim_{t\to +\infty}} A^U_2 = {\displaystyle \lim_{t\to +\infty}} A^U_3 = 0$.
		\hfil\hfil $\blacksquare$
	\end{proposition}
	Proposition \ref{th14} says that the growth of the density of alleles $r$ pertaining to uninfected exceeds that of the density of alleles $s$ pertaining to uninfected.
	This property is insensitive to the introduction of infected mosquitoes, including say ``massive'' release of infected homozygous with genotype $(s,s)$, and derives from the involved mechanisms of genetic transmission.
	See related results in \cite[Lemmas 16, 17, 18]{Perez:2020aa}.
	
	\begin{figure*}[h]
		\begin{gather}
		\label{eq68}
		v_{A1}(t)
		= \left|
		\tm^U(t) \frac{|A^U|}{|A^U|+|A^W|} - \tmu^U(t) - \left(
		\tm^W(t) - \tmu^W(t)
		\right) +\varepsilon
		\right|_+ |A^W|,\ v_{A2}(t)=v_{A3}(t)=0\\
		\label{eq68bis}
		v_{A1}(t)=v_{A2}(t)=0,\
		v_{A3}(t)
		= \left|
		\tm^U(t) \frac{|A^U|}{|A^U|+|A^W|} - \tmu^U(t) - \left(
		\tm^W(t) - \tmu^W(t)
		\right) +\varepsilon
		\right|_+ |A^W|
		\end{gather}
		\hrule
	\end{figure*}
	
	{\em Hint of proof:}
	$\bullet$
	From \eqref{eq1118a} one shows that $\frac{d|A^U|}{dt} \leq (m^U_1(b^*(\alpha(A)))-\mu^U_1) |A^U|$, which is negative for large values of $|A|$, so that $|A^U|$ is uniformly bounded on $[0,+\infty)$.
	$\bullet$
	Consider a trajectory for which initially $A^U_i(0)>0$, for some $i\in\{1,2,3\}$.
	Due to Theorem \ref{th15}, 
	$A^U_i(t)>0$, $t\geq 0$.
	From \eqref{eq5102}-\eqref{eq620}, one gets
	$\frac{d}{dt} \left[
	\ln\left(
	\frac{A^U_r}{A^U_s}
	\right)\right]
	> 0$,
	so that the ratio $\frac{A^U_s}{A^U_r}$ decreases along time.
	One then shows that, for {\em each} trajectory, exist $c_1,c_2>0$ so that
	$c_1 A^U_1(t) \geq A^U_2(t) \geq c_2 A^U_3(t)$, $t\geq 0$.
	Invoking uniform boundedness of the trajectories and Proposition \ref{th11} yields
	$\lim\limits_{t\to +\infty}\frac{A^U_s(t)}{A^U_r(t)}=0$.
	\hfil\hfil $\square$
	
	\subsection{State-feedback control laws and stabilisation results}
	
	We now present the main results.
	The principle of the stabilization method is to extend ideas from \cite{Bliman:2020aa} to the representation \eqref{eq1118},
	with maps $\tm^U, \tmu^U, \tm^W, \tmu^W$ defined in \eqref{eq16-117}.
	The first result concerns release of {\em resistant} mosquitoes.
	
	\begin{theorem}[Infection by release of resistant infected]
		\label{th20}
		Let $b^{**}:= b^*(A^{W*}_r(e^W\otimes e_1))$ be the population level corresponding to the resistant {\em Wolbachia} infected homozygote.
Assume $\mu^U_1+m^W_2(b^{**}) -\mu^W_2 > 0$; $v_{A1}\not\equiv 0$; and \eqref{eq68} holds for any large enough $t$, for some $\varepsilon \in (0,\mu^U_1+m^W_2(b^{**})-\mu^W_2)\cap (0,\mu^U_1)$.
		Then for any solution of \eqref{eq0},
		${\displaystyle \lim_{t\to +\infty}} A(t) = A^{W*}_r (e^W\otimes e_1)$,
		and there exists $T\geq 0$, such that $v_A\equiv 0 \text{ on } [T,+\infty)$.
		\hfil\hfil $\blacksquare$
	\end{theorem}
	
	Choosing control \eqref{eq68} thus allows to reach full infection by use of control vanishing in finite time.
	Theorem \ref{th20bis} is analogous, with {\em susceptible} mosquitoes.
	Of course, releasing susceptible or resistant requires different quantities of insects to achieve infection, see numerical essays in \cite{Perez:2020ab}.
	
	\begin{theorem}[Infection by release of susceptible infected]
		\label{th20bis}
		Assume the setting of Theorem {\rm\ref{th20}} holds with $\mu^U_1+m^W_3(b^{**}) -\mu^W_3 > 0$, $\varepsilon \in (0,\mu^U_1+m^W_3(b^{**})-\mu^W_3)\cap (0,\mu^U_1)$, and \eqref{eq68bis} instead of \eqref{eq68}.
		Then the same conclusions hold.
		\hfil\hfil $\blacksquare$
	\end{theorem}
	
	When \eqref{eq68} or \eqref{eq68bis} applies,
	then \eqref{eq1118} yields
	$\frac{1}{|A^W|}\frac{d|A^W|}{dt}=$ 
	$\max\left\{
	\frac{1}{|A^U|}\frac{d|A^U|}{dt} +\varepsilon, \tm^W(t) - \tmu^W(t)
	\right\}$.
	The mean growth rate of {\em Wolbachia} infected is thus kept unchanged when larger than the mean growth rate of uninfected plus $\varepsilon$; and changed to this value otherwise.
	The feedback law thus ensures that {\em the mean growth rate of infected mosquitoes is always  larger than the mean growth rate of uninfected.}
	This is  the principle of the proposed stabilization method.
	
	
	{\em Hint of the proofs:}
	The proof of Theorems \ref{th20} and \ref{th20bis} are similar, and based on the following successive steps.
	First prove that the proposed (linear in state) control yields (uniformly ultimately) bounded trajectories.
	Using Proposition \ref{th11}, this shows that the uninfected population vanishes asymptotically, as well as the ratio between resistant uninfected and resistant infected population.
	Using the inequalities assumed in the statements, this implies that the control feedback term in \eqref{eq68} or \eqref{eq68bis} is zero from a certain time and beyond.
	The system then behaves asymptotically as a mixing of {\em infected only} mosquitoes of different genotypes, and by virtue of Proposition \ref{th14} converges towards the equilibrium with the best fitness, {\em i.e.}\ the resistant monomorphic equilibrium.
	\hfil\hfil $\square$
	
	\section{Numerical simulations}
	\label{se6}
	
\begin{figure*}[h]
		\begin{subfigure}[t]{0.40\textwidth}
			\includegraphics[width=\textwidth]{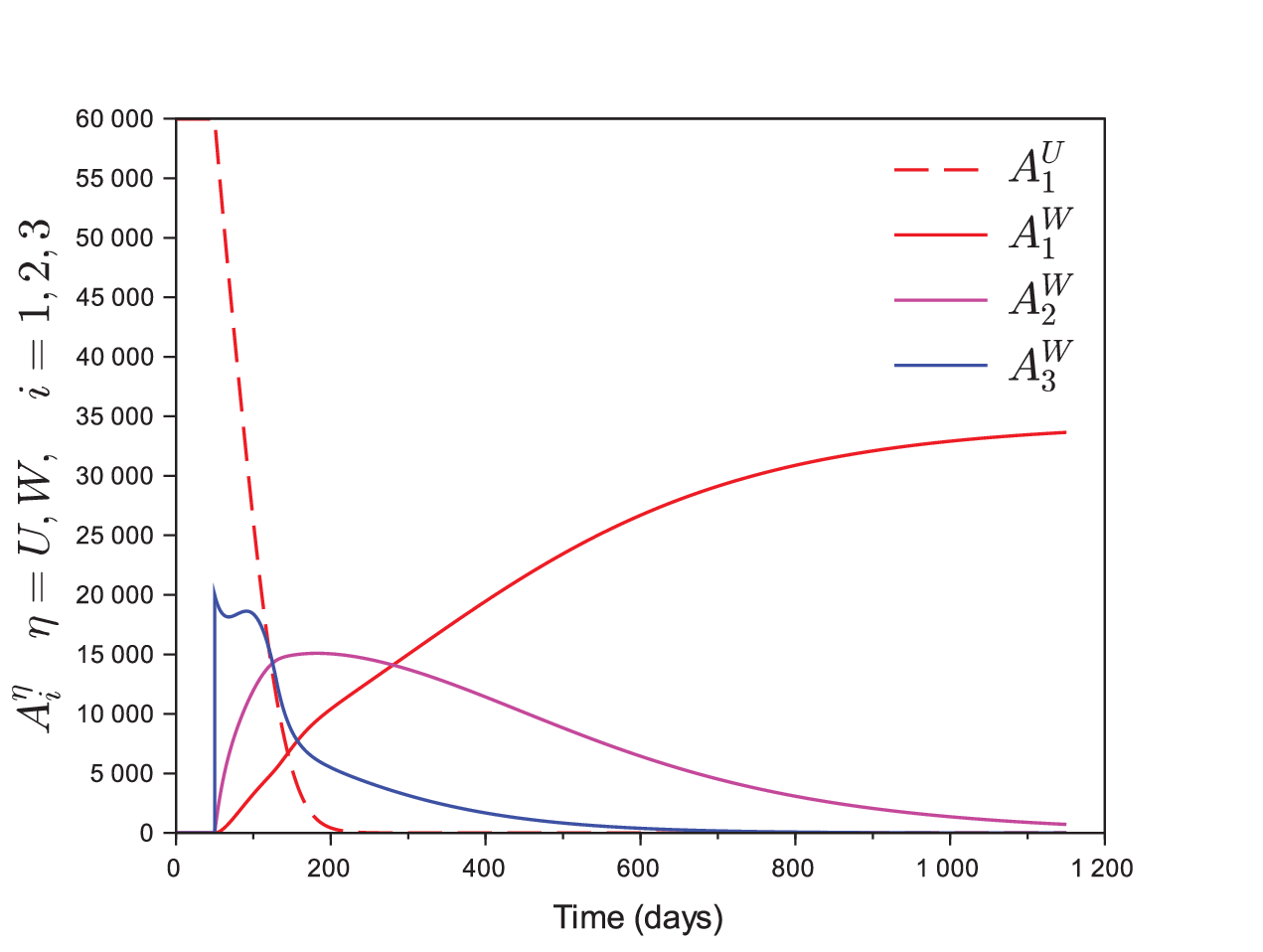}
			\caption{Solution of the controlled system}
			\label{fig:I101}%
		\end{subfigure}
		\hfill
		\begin{subfigure}[t]{0.40\textwidth}
			\includegraphics[width=\textwidth]{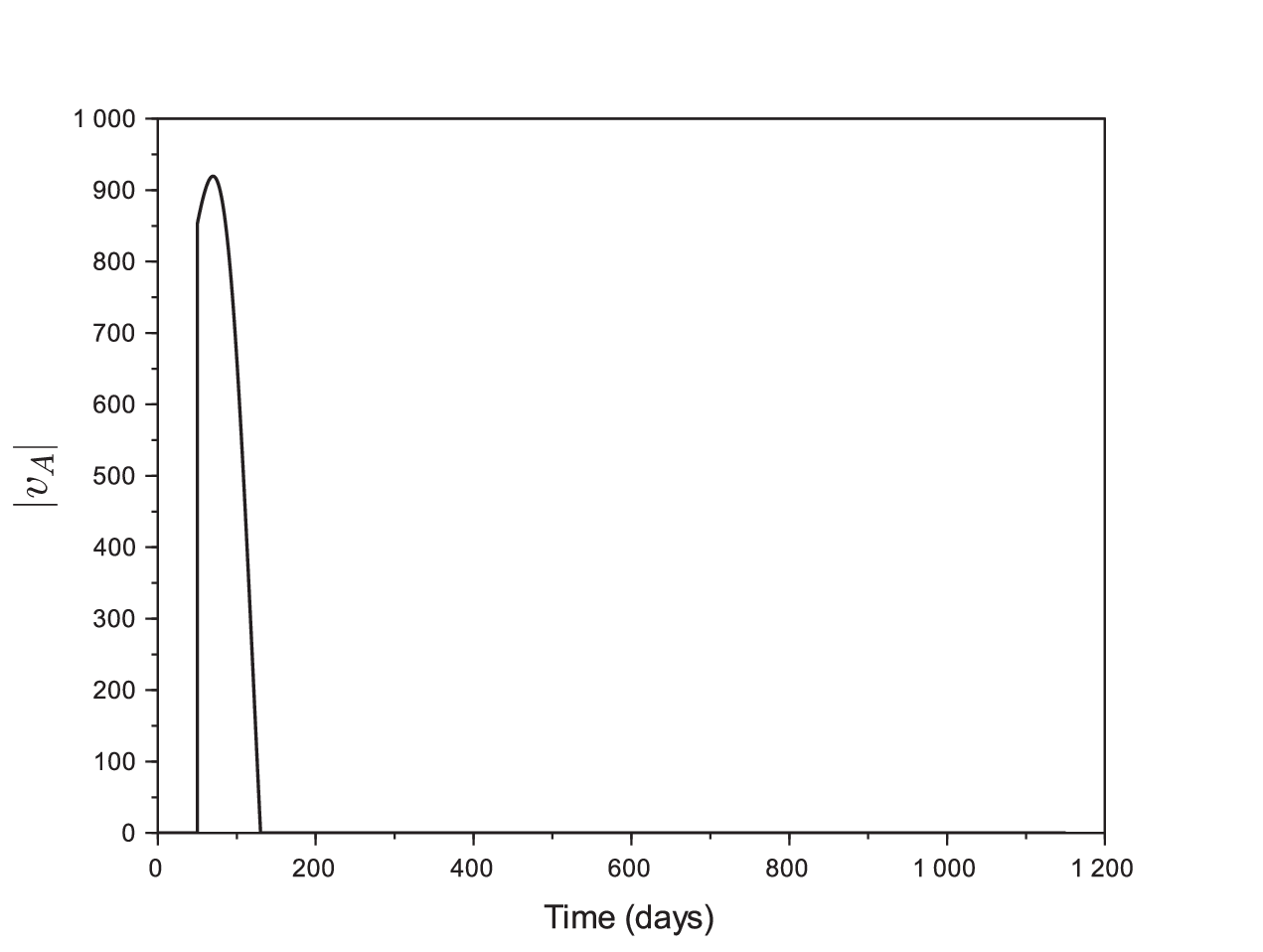}
			\caption{Control effort $v_{A3}(t)$}
			\label{fig:I102}%
		\end{subfigure}
		\hfill
		\begin{subfigure}[t]{0.40\textwidth}
			\includegraphics[width=\textwidth]{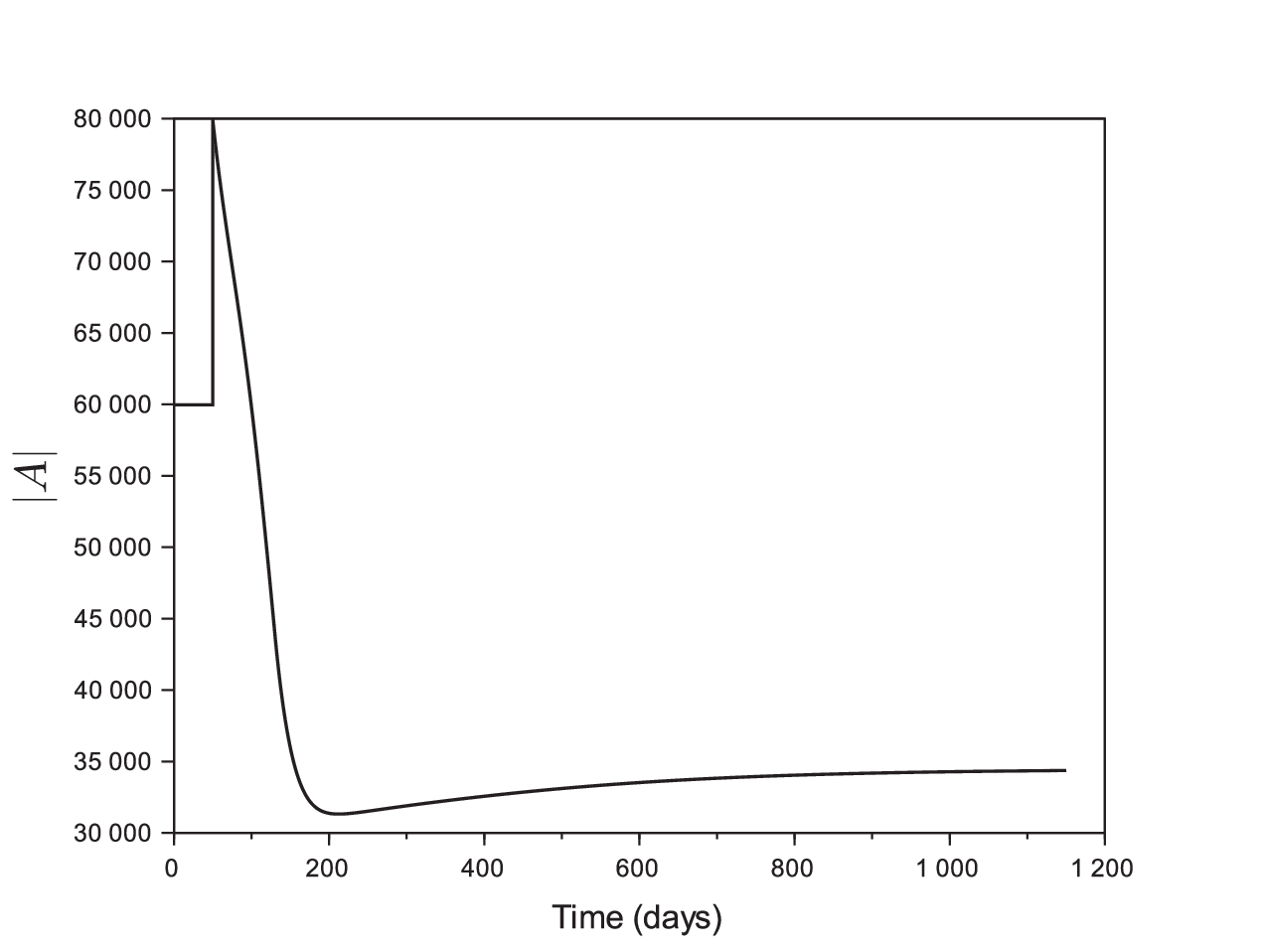}
			\caption{Total population $|A(t)|$}
			\label{fig:I103}%
		\end{subfigure}
		\hfill
		\begin{subfigure}[t]{0.40\textwidth}
			\includegraphics[width=\textwidth]{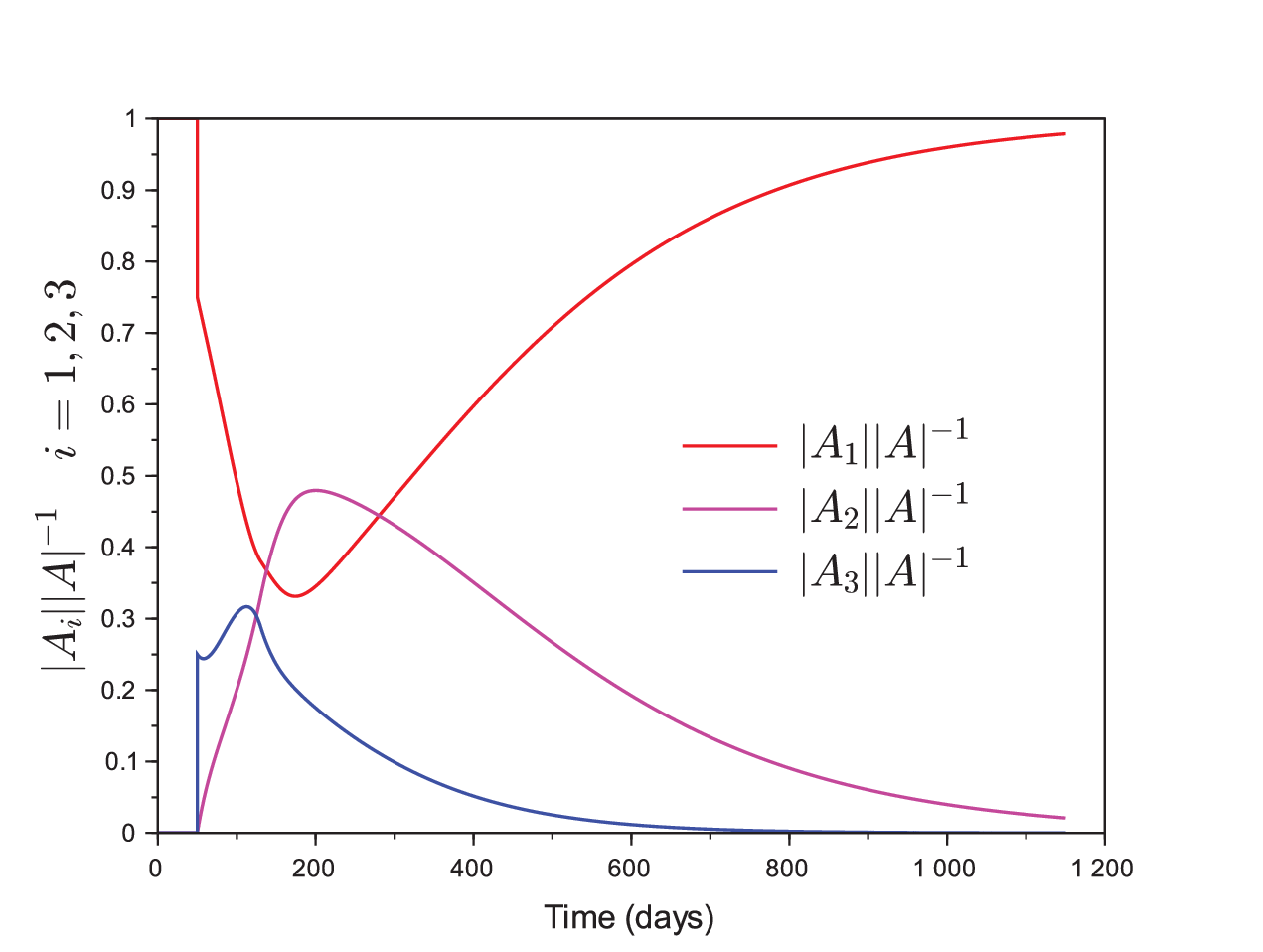}
			\caption{Genotype frequencies $\frac{|A_i(t)|}{|A(t)|}$}
			\label{fig:I104}%
		\end{subfigure}
		\hfill
		\begin{subfigure}[t]{0.40\textwidth}
			\includegraphics[width=\textwidth]{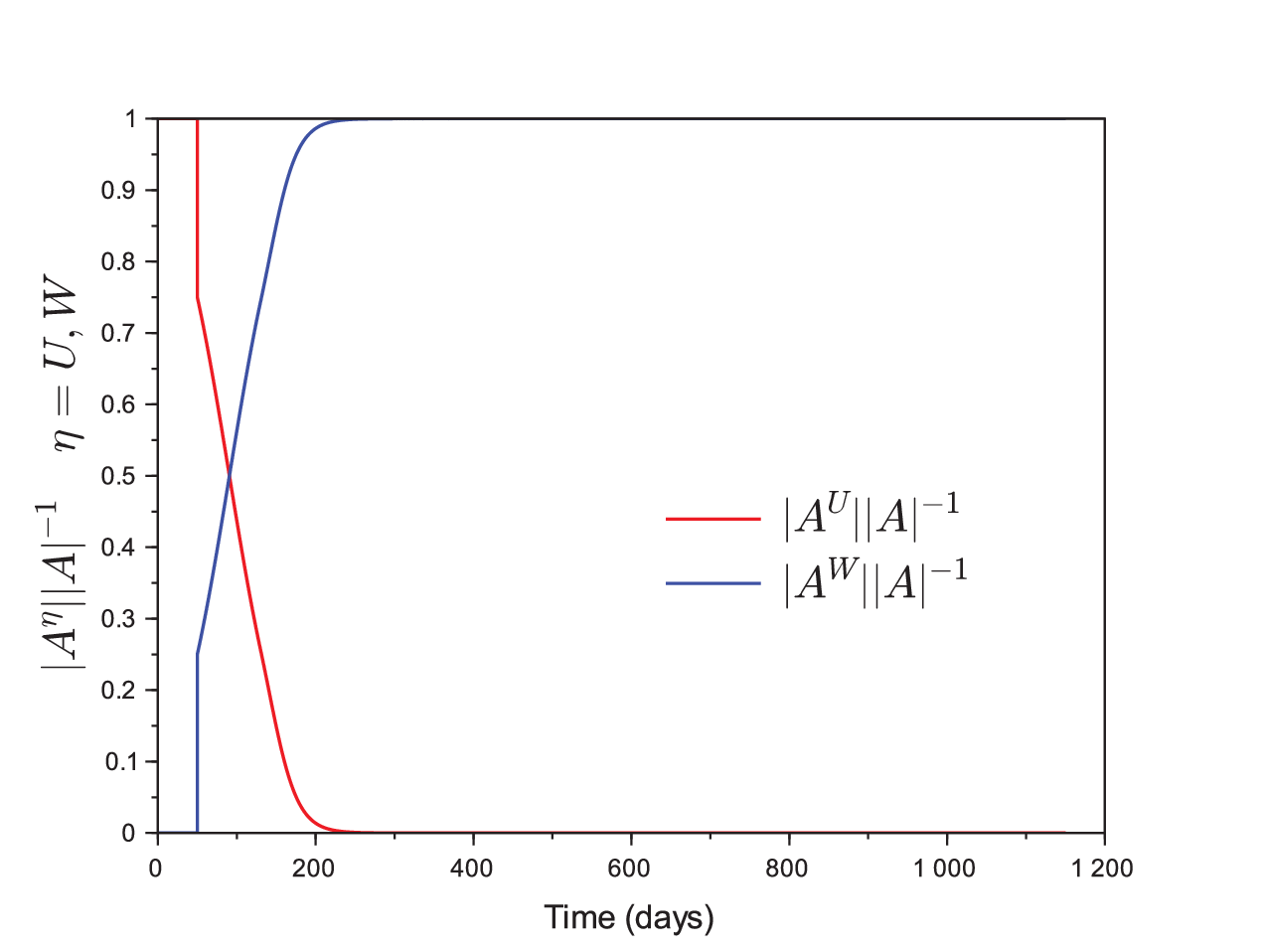}
			\caption{Relative frequencies $\frac{|A^\eta(t)|}{|A(t)|}$ of uninfected/infected}
			\label{fig:I105}%
		\end{subfigure}
		\hfill
		\begin{subfigure}[t]{0.40\textwidth}
			\includegraphics[width=\textwidth]{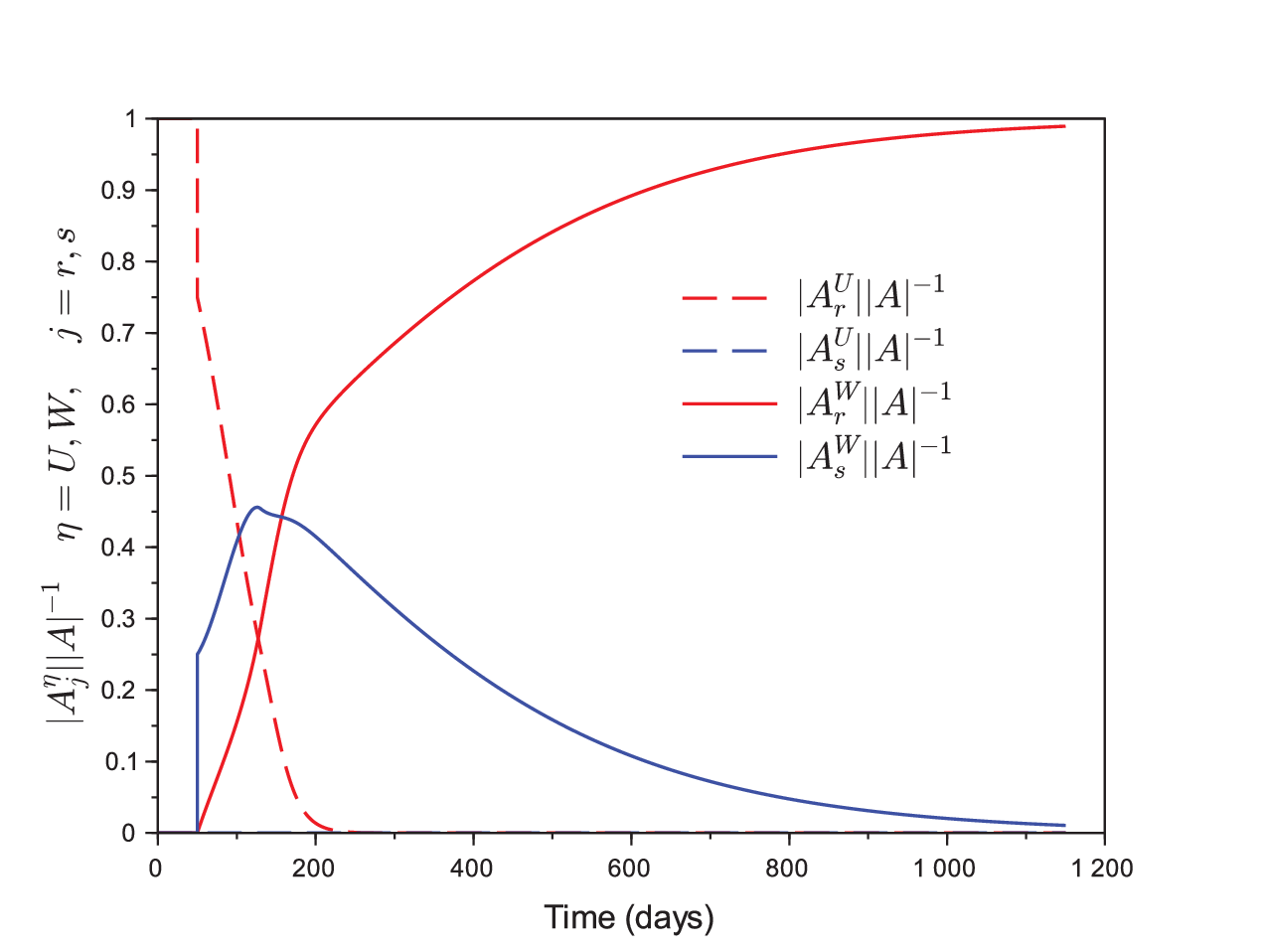}
			\caption{Allelic frequencies of uninfected/infected $\frac{|A^\eta_j(t)|}{|A(t)|}$}
			\label{fig:I106}%
		\end{subfigure}
		\caption{Release of insecticide-susceptible mosquitoes infected by {\em Wolbachia} in a resistant population, according to \eqref{eq68bis}.}
		\label{fig:S.I}
		\hrule
	\end{figure*}

	Release of homozygous insecticide-susceptible {\em Wolbachia}-infected mosquitoes in an environment subject to adulticide and larvicide by applying control \eqref{eq68bis} is shown in Fig.\ \ref{fig:S.I}.
	We assume 5\% of relative increase in mortality of infected larvae/adults, 7\% (resp.\ 3.5\%) of relative mortality decrease to resistant homozygote (resp.\ to heterozygote).
	Parameters are taken from \cite{Walker:2011aa,Hoffmann:2014aa,Koiller:2014aa,Adekunle:2019aa,Xue:2017aa,Bliman:2018aa,Styer:2007aa,McMeniman:2009aa,McMeniman:2010aa,Luz:2009aa}
	(all units in days$^{-1}$): $ \hmu_i^\eta(b)  + \nu =  \hmu_{i0}^\eta(1+\hat{\mu} b)$, $\nu=1/10$, $\hmu_{10}^U=0.093$, 
$\hmu_{20}^U=0.097$,
$\hmu_{30}^U= 0.1$,
$\hmu_{10}^W=0.098$,
$\hmu_{20}^W=\hmu_{30}^W=0.105$,
$\hmu=0.01$,
$\mu_{1}^U= 0.057$,
$\mu_{2}^U=0.059$,
$\mu_{3}^U=0.061$,
$\mu_{1}^W=0.060$, $\mu_{2}^W=0.062$, $\mu_{3}^W=0.064$,
$\omega^U=18$, $\omega^W=12$.
Initial condition is at monomorphic resistant uninfected equilibrium, $A^{U*}_r = \frac{\nu}{\hmu\mu_{i}^\eta}\left(\frac{\nu \omega^U}{\hmu_{i0}^U\mu_{i}^U}-1 \right) \simeq \num{59963.805}$.
	An impulse of susceptible infected mosquitoes equal to $\frac{1}{3}A^{U*}_r$
	is released at $t=50$ days.
	The total amount of released mosquitoes is $\num{345133.30}$, about 5.75 times the initial value $A^{U*}_r$.
	
	\section{Conclusion}
	\label{se7}
	
	We proposed a two-life phase model accounting for Mendelian and maternal inheritance, allowing to consider chemical and biological vector control in a unified framework.
	Feedback laws have been proposed and shown to induce {\em Wolbachia} infection in any situation.
	To deal with the lack of full state measurement and the non-permanent nature of the releases, future research will study output stabilization and impulsive control.
	Also, reducing the total number of released mosquitoes by use of insecticide will be considered.
	
	\setcounter{section}{0}%
	\setcounter{subsection}{0}%
	\setcounter{equation}{0}%
	\setcounter{theorem}{0}%
	\def\thesection{A.\arabic{section}}
	\def\thetheorem{A.\arabic{theorem}}
	\def\thelemma{A.\arabic{lemma}}
	\def\theequation {A.\arabic{equation}}
	
	\section*{Appendix - The heredity matrix function $\alpha$}
	
	By convention, $i=1$, resp.\ $i=3$, if $j=r$, resp.\ $j=s$.
	
	\begin{lemma}
		\label{le10}
		For any $\eta=U,W$, $i=1,2,3$, $\alpha(e^\eta \otimes e_i) = (e^\eta \otimes e_i)$.
		\hfil\hfil $\blacksquare$
	\end{lemma}
	
	\begin{lemma}
		\label{le1}
		For any $A\in\Rset_+^6$ and any $j=r,s$,
		\begin{gather*}
		\label{eq997y}
		\alpha^U_i (A) + \frac{1}{2} \alpha^U_2(A)
		= \frac{|A^U|}{|A|} A^U_j,\\
		\label{eq997z}
		\alpha^W_i (A) + \frac{1}{2} \alpha^W_2(A)
		= \frac{1}{2} A^W_j + \frac{1}{2} \frac{|A_j| |A^W|}{|A|},\\
		\label{eq997a}
		\sum_{i=1}^3 \alpha^U_i (A)
		= \frac{|A^U|^2}{|A|},\qquad
		\sum_{i=1}^3 \alpha^W_i (A)
		= |A^W|,\\
		\label{eq997b}
		|\alpha(A)| = \sum_{\eta=U,W}\sum_{i=1}^3 \alpha^\eta_i (A)
		= |A| - \frac{1}{|A|} |A^U| |A^W|,\\
		\label{eq997c}
		\sum_{\eta=U,W} \alpha^\eta_i (A)
		=  \frac{1}{|A|} \left(
		|A_j|^2 - A^W_j A^U_j
		\right),\\
		\label{eq997d}
		\sum_{\eta=U,W} \alpha^\eta_2 (A)
		=  \frac{1}{|A|} \left(
		2|A_r| |A_s| - A^W_r A^U_s - A^W_s A^U_r
		\right).
		\hspace{.3cm} \blacksquare
		\end{gather*}
	\end{lemma}
	
	\begin{lemma}
		\label{le34}
		Let $A\in\Rset_+^6$.
		\begin{enumerate}
			\item
			\label{le34i}
			For any $j=r,s$,
			$A_j = 0_2$ implies $\alpha^\eta_i(A) = \alpha^\eta_2(A) = 0$, $\eta=U,W$.
			\item
			\label{le34ii}
			For any $\eta=U,W$, $A^\eta = 0_3$ yields $\alpha^\eta_i(A) = 0$, $i=1,2,3$.
			\item
			\label{le34iii}
			For any $\eta=U,W$ and $i,i'=1,2,3$ such that $\{i,i'\}=\{1,3\}$ or $2\in\{i,i'\}$,
			$A^\eta_i>0, A^\eta_{i'}>0$ yields $\alpha^\eta_{i''}(A)>0$, $i''=1,2,3$.
			\item
			\label{le34iv}
			For any $i,i'=1,2,3$ such that $\{i,i'\}=\{1,3\}$ or $2\in\{i,i'\}$, $A^U_i>0, A^W_{i'}>0$ implies
			$\alpha^W_{i''}(A)>0$ or $\alpha^W_{i''}(\alpha(A))>0$, $i''=1,2,3$.
			\item
			\label{le34v}
			For any $j=r,s$, $A^U_j=0\Rightarrow \alpha^U_i(A) = \alpha^U_2(A) = 0$.
			\hfil\hfil $\blacksquare$
		\end{enumerate}
	\end{lemma}
	
	Points \ref{le34i}, \ref{le34ii} indicate that in monomorphic state, all offsprings have identical homozygous genotype; and in homogeneous uninfected (resp.\ infected) state, all offsprings are uninfected (resp.\ infected).
	The other points describe the result of mixing.
	Point \ref{le34iii} states that if two different uninfected (resp.\ infected) genotypes are present in some state, or if heterozygous are present, then the birth rate of every uninfected (resp.\ infected) genotype is positive: 
	both alleles are present, and all genotypes are thus present in the offspring.
	More intricate, point \ref{le34iv} says that if a genotype is present in uninfected mosquitoes and the other one in infected, or if an heterozygous is present in a heterogeneous state, then the birth rate of every {\em infected} genotype is positive.
	Appearance of missing genotypes occurs ``directly" in point \ref{le34iii} during 1st mating ($ \alpha^\eta_{i''}(A)>0$), but this may happen ``indirectly", after a 2nd mating:  $\alpha^\eta_{i''}(\alpha(A))>0$.
	For example, when mixing uninfected of genotype $(r,r)$ with infected of genotype $(s,s)$, infected of genotypes $(s,s)$ and $(r,s)$ arise from first mating, and of genotype $(r,r)$ only after second one.
	Due to complete CI, the symmetric property is {\em not} true for the uninfected birth rate, as  {\em e.g.}\ mixing of uninfected mosquitoes bearing genotype $(r,r)$ with infected of any genotype only produces uninfected of identical genotype. This property is the meaning of point \ref{le34v}.
	
	\bibliographystyle{IEEEtran}
	\bibliography{ecc21-bib}

	

\end{document}